\documentstyle[11pt,aaspp4,epsf]{article}

\newcommand{\simgt}{\lower.5ex\hbox{$\; \buildrel > \over \sim \;$}}
\newcommand{\simlt}{\lower.5ex\hbox{$\; \buildrel < \over \sim \;$}}

\newcommand{\cpara}{c_{\scriptscriptstyle \|}}
\newcommand{\cperp}{c_{\scriptscriptstyle \bot}}
\def\P{L}
\def\wq{{w_Q}}
\def\bfk{{\bf k}}

\def\callr{{\cal R}}

\newcommand{\bfqpara}{{ q}_{\scriptscriptstyle \|}}
\newcommand{\bfqperp}{{\bf q}_{\scriptscriptstyle \bot}}

\begin{document}

\title{Can Geometric Test Probe the Cosmic Equation of State ?} 

\bigskip
\author{Kazuhiro {\small YAMAMOTO} and Hiroaki {\small NISHIOKA}
\\
{\it Department of Physics, Hiroshima
     University, Higashi-Hiroshima 739-8526, Japan}}

\begin{abstract}
{}Feasibility of the geometric test as a probe of the cosmic equation 
of state of the dark energy is discussed assuming the future 2dF QSO
sample. We examine sensitivity of the QSO two-point correlation 
functions, which are theoretically computed incorporating the 
light-cone effect and the redshift distortions, as well as the 
nonlinear effect, to a bias model whose evolution is 
phenomenologically parameterized. It is shown that the correlation 
functions are sensitive on a mean amplitude of the bias and not to 
the speed of the redshift evolution.
We will also demonstrate that an optimistic geometric test could 
suffer from confusion that a signal from the cosmological model 
can be confused with that from a stochastic character of the bias. 
\end{abstract}


\def\WQ{w_{\rm Q}}
\section{INTRODUCTION}
Recent observations of the cosmic microwave background anisotropies
and the distant type Ia supernovae favor a spatially flat universe 
whose expansion is currently accelerating (de Bernardis et al.~2000; 
Lange et al.~2000; 
Perlmutter et al.~1999a; Riess et al.~1998). Motivated by this fact, 
variants of the cold dark matter (CDM) model with the cosmological 
constant ($\Lambda$CDM) have been widely studied. In particular, 
quintessence is proposed as a plausible model (e.g., Caldwell et 
al.~1998; Zlatev et al.~1998). 
An attractive feature of the quintessence is that certain 
quintessence models (tracker models) naturally explain the 
'coincidence problem', the near coincidence of the density of 
matter and the dark energy component at present (Zlatev et al.~1998; 
Steinhardt et al.~1999, and references therein).
Observational constraints have been investigated on the quintessential 
cold dark matter (QCDM) model (e.g., Efstathiou 1999; Perlmutter et 
al.~1999b; Wang et al.~2000; Newman \& Davis 2000), including attempts 
for reconstructing the cosmic equation of state or the quintessential 
potential (Starobinsky 1998; Saihi et al.~2000; Nakamura \& Chiba 1999; 
Chiba \& Nakamura 2000).

In the meanwhile, recent progress on the 2dF QSO redshift (2QZ) survey 
has been reported by Shanks et al.~(2000). 
An interesting scientific aim of the 2QZ survey is to obtain 
new constraints on the cosmological constant from the geometric 
test, which was originally proposed by Alcock \& Pancynski (1979), 
with the QSO clustering statistics. 
Several authors have proposed possible cosmological tests using the 
clustering of high-redshift objects (Ryden 1995; 
Ballinger et al.~1996; Matsubara \& Suto 1996; Popowski et al.~1998; 
Nair 1999).
In a series of work (Suto et al.~2000; Yamamoto et al.~2000, 
hereafter Paper I; references therein), 
a useful theoretical formula has been developed for predicting 
the correlation functions of high-$z$ objects, incorporating 
the light-cone effect and the redshift-space distortions, as 
well as the geometric distortion, simaltaneously. 
The theoretical formula is useful because it is expressed in a 
semi-analytic form, which enable us to compute the two-point 
correlation functions corresponding to a survey sample based on 
specific cosmological models without huge numerical simulations. 
Therefore it will be worth to examine the feasibility of the 
geometric test as a probe of the cosmic equation of state of 
the dark energy with assuming a realistic QSO clustering statistics.

Unfortunately, however, the QSO clustering bias is not well-understood
at present. This ambiguity of the clustering bias should 
be crucial for the geometric test because the evolution of bias 
affects the predicted correlation functions. In Paper I, behavior 
of the QSO correlation functions is partially examined.
On the other hand the stochastic bias has been discussed as a possible 
character of the galaxy biasing by several authors,
(e.g., Dekel \& Lahav 1999; Tegmark \& Peebles 1998; Taruya et al.~1998). 
If the QSO clustering bias possesses the stochastic character, 
it affects the redshift-space correlation functions (Pen 1998).
In the present paper we consider the feasibility of the geometric 
test as a probe of the cosmic equation of state of the dark 
energy component, including the possible stochasticity of the 
QSO clustering bias. 
This paper is organized as follows. In section 2, we briefly review 
the theoretical formula for the two-point statistics. In section 3, 
sensitivity of the correlation functions to the evolution of a bias 
model is investigated. Section 4 is devoted to summary and conclusions. 
Throughout this paper we use the unit in which the light velocity 
$c$ equals $1$.

\section{Contents for Theoretical Prediction}
We restrict ourselves to a spatially flat FRW universe, 
and follow the quintessential cosmological model consisting
a scalar field slowly rolling down its effective potential. 
The effective equation of state of the dark energy, $\wq=p_Q/\rho_Q$,
can be a function of redshift in general case, however, we will 
consider the QCDM model with a constant equation of state for 
simplicity. In this case the dark energy 
density evolves $\rho_Q\propto a(z)^{-3(1+\wq)}$, where $a(z)$ is 
the scale factor. Then the relation between 
the comoving distance and the redshift is 
\begin{eqnarray}
  r(z)={1\over H_0}\int_0^z{dz'\over
  [\Omega_m(1+z')^3+\Omega_Q(1+z')^{3(1+\wq)}]^{1/2}},
\label{defr}
\end{eqnarray}
where $H_0=100h{\rm km/s/Mpc}$ is the Hubble constant,
$\Omega_m$ and $\Omega_Q(=1-\Omega_m)$ denote the density 
parameters of the matter component and the dark energy, 
respectively, at present.

Wang \& Steinhardt (1998) have given a useful approximate formula
for the linear growth index, which we adopt for computation of
the $f$-factor defined by $f(z)\equiv{d \ln D_1(z) / d\ln a(z)}$,
where $D_1(z)$ is the linear growth rate.
Simple fitting formulas for the linear transfer function and 
the nonlinear mass perturbation power spectrum are presented 
by Ma et al.~(1999) for the QCDM model, which we adopt for
modeling the CDM mass density perturbations.
The formulas are applicable for the QCDM model with the constant 
equation of state $-1\leq \wq \simlt -1/3$ and up to $z\sim 4$
with errors $\simlt$ 10 \%.
In addition we adopt the normalization by cluster abundance 
(Wang \& Steinhardt 1998; Wang et al.~2000).
Throughout this paper we assume the Harrison-Zeldovich spectrum. 

In predicting clustering statistics of high-redshift objects in a redshift 
survey, several observational effects must be incorporated for careful 
comparison between theoretical predictions and observational results. 
A useful theoretical formula for the two-point statistics has been 
developed incorporating the redshift distortions due to peculiar 
motion of sources and the  light-cone effect simaltaneously, 
as well as the geometric distortion (Suto et al.~2000; Paper I). 
According to the theoretical formula, the two-point correlation function 
is given by a Fourier transform of the power spectrum on a light cone 
$P^{\rm LC}_l(k)$
\begin{equation}
  \xi_l(R)={1\over2\pi^2 i^l}\int_0^\infty dk k^2 j_l(kR) P^{\rm LC}_l(k),
\label{Fourie}
\end{equation}
where $P^{\rm LC}_l(k)$ is obtained by averaging the local power 
spectrum $P_l^{\rm crs}(k,z)$ over the redshift
\begin{equation}
  P^{\rm LC}_l(k)={{\int dz W(z) P_l^{\rm crs}(k,z)}
             \over {\int dz W(z)}}
\label{Pcllk}
\end{equation}
with the weight factor 
\begin{equation}
  W(z)=\biggl({dN\over dz}\biggr)^2\biggl({s^2 ds\over dz}\biggr)^{-1},
\label{weight}
\end{equation}
where $dN/dz$ denotes the number count of the objects per unit redshift 
and per unit solid angle, and $s=s(z)$ denotes the distance-redshift 
relation of the radial coordinate that we chose to plot a map of sources. 
In the present paper we adopt the distance-redshift relation of the 
Einstein de Sitter universe, i.e., $s(z)={2 H_0^{-1}}[1-{(1+z)^{-1/2}}]$.
In equation (\ref{Pcllk}) $z$-integration arises from the light-cone 
effect within the small-angle approximation.
The power spectrum $P_l^{\rm crs}(k,z)$ in (\ref{Pcllk}) is given by 
\begin{eqnarray}
  &&P_l^{\rm crs}(k,z)={2l+1\over \cperp^2\cpara}\int_0^1 d\mu 
  \P_l(\mu) 
\nonumber
\\
  && \hspace{1cm}\times 
  P_{\rm QSO}\biggl(\bfqpara\rightarrow{k\mu\over\cpara},~
  |\bfqperp|\rightarrow{k\sqrt{1-\mu^2}\over\cperp},z\biggr),
\label{scaling}
\end{eqnarray}
where $P_{\rm QSO}(\bfqpara,|\bfqperp|,z)$ is the QSO power spectrum,
$\bfqpara$ ($\bfqperp$) is the wave number component parallel 
(perpendicular) to the line-of-sight direction in the real space, 
and $\P_l(\mu)$ is the Legendre polynomial.
In equation (\ref{scaling}), we denoted $k=|\bfk|$, 
$\cperp={r(z)/s(z)}$ and $\cpara={dr(z)/ds(z)}$ with
the comoving distance in the real space $r(z)$, 
equation (\ref{defr}).

We model the power spectrum of QSO distribution 
by introducing the bias factor $b(z)$,
\begin{eqnarray}
  &&P_{\rm QSO}(\bfqpara,|\bfqperp|,z)=
  \biggl\{b(z)^2+2b(z)f(z)\callr(z)\biggl({\bfqpara\over q}\biggr)^2
\nonumber
\\
  &&\hspace{1.5cm}
  +f(z)^2\biggl({\bfqpara\over q}\biggr)^4\biggr\}
  P_{\rm mass}(q,z)D\Bigl[\bfqpara \sigma_P(z)\Bigr],
\label{appendixP}
\end{eqnarray}
where $q=\sqrt{\bfqpara^2+|\bfqperp|^2}$, and $P_{\rm mass}(q,z)$ 
is the CDM mass power spectrum.
The terms in proportion to $f(z)$ in (\ref{appendixP})
is traced back to the linear distortion (Kaiser) effect. 
To describe a cross correlation between the QSO distribution and 
the CDM mass distribution, we here introduced the cross correlation
coefficient $\callr(z)$. 
In the case of the deterministic bias, the cross correlation
coefficient can be set $\callr(z)=1$, however, in the case of the
stochastic bias, the cross correlation coefficient is allowed
to deviate from unity (e.g., Pen 1998).  

In (\ref{appendixP}), $D[\bfqpara \sigma_P(z)]$ is the damping factor 
due to the Finger-of-God effect, for which we adopt the exponential 
model for the distribution of the pairwise velocity dispersion 
$\sigma_P$. For $\sigma_P$ we adopt an approximate formula whose 
validity is investigated by several authors 
(see Magira et al.~2000; Mo et al.~1997). 
The nonlinear effect is not a dominant effect on large length scales, 
however, it can not be neglected. Therefore we here take the 
nonlinear effects into account for definiteness.

\def\rmB{{\rm B}}
Recently, Boyle et al.~(2000) have reported on the evolution of QSO 
optical luminosity function using their preliminary result of the 2QZ 
survey. The evolution of the luminosity function is compared with
analytic fitting formulas. For theoretical predictions which
correspond to the on-going 2QZ survey, we adopt their quasar 
luminosity function with best-fitted parameters of a power-low 
polynomial evolution of luminosity under the assumption of the 
Einstein de Sitter universe.  
Then the number count $dN/dz$ brighter than the limiting
magnitude $B\leq 20.85$ can be obtained by integrating 
the luminosity function. 
\footnote{For the K-correction we assume the quasar energy spectrum 
$L_\nu\propto\nu^{-0.5}$.}

\section{Sensitivity}
The QSO clustering bias is a challenging problem and a few authors
have proposed theoretical models (Fang \& Jing 1998; Martini \& 
Weinberg 2000; Haiman \& Hui 2000). However, these models seem
to be prototype models, hence we here adopt a model whose evolution
is phenomenologically parameterized, for simplicity, 
as (c.f., Matarrese et al.~1997),
\begin{equation}
  b(z)=\alpha+\bigl(b(z_*)-\alpha\bigr)
  \biggl({1+z\over 1+z_*}\biggr)^\beta,
\label{defbz}
\end{equation}
where $\alpha$, $\beta$ and $b(z_*)$ are free parameters, 
$\beta$ specifies the speed of redshift evolution,
$b(z_*)$ is the amplitude at a mean redshift $z_*$, 
and $\alpha$ corresponds to the amplitude of bias at $z=0$
in the limit of $z_*\gg1$. Throughout the present paper we 
fix $\alpha=0.5$, however, this does not change the results
for $0\simlt \alpha\simlt 1$. We define the mean redshift by
\begin{eqnarray}
  z_*={{\int_{z_{\rm min}}^{z_{\rm max}}dz z W(z)} 
  \over{\int_{z_{\rm min}}^{z_{\rm max}}dz W(z)}},
\label{zstar}
\end{eqnarray}
with the weight factor (\ref{weight}). Here we consider the
sample in the range $0.3\leq z\leq 2.2$, in this case we have
$z_*=1.2$. We note that $b(z_*)$ is almost same as the mean 
amplitude of the bias defined in the similar way to (\ref{zstar}).

{}For the geometric test, the ratio of the correlation functions
$\xi_2(R)/\xi_0(R)$ will be an important quantity to characterize 
the geometric distortion effect. 
{}Figure 1 shows contours of $\xi_2/\xi_0$ on the 
$(\beta-b(z_*))$ plane with the separation $R$ fixed as 
$20h^{-1}{\rm Mpc}$ for various cosmological models ({\it solid lines}).
Panels (a) and (b) show  the cases of the $\Lambda$CDM model 
$(\Omega_m=0.3,~\wq=-1)$ and the QCDM model 
$(\Omega_m=0.3,~\wq=-1/2)$, respectively. In both panels we show
the case of the deterministic bias. 
This figure shows that the value of $\xi_2/\xi_0$ is sensitive 
only to the parameter $b(z_*)$, and not to the speed of evolution $\beta$. 

Recently it is reported that the QSO correlation 
function is consistent with being $\xi=(r/r_0)^{-1.8}$ with 
$r_0=4~{h^{-1}\rm Mpc}$ from the preliminary result of the 2QZ survey 
(Shanks et al.~2000). Previous result of the QSO surveys similar 
to the 2QZ survey reported $r_0=6~{h^{-1}\rm Mpc}$ 
(Croom \& Shanks 1996). The dashed lines on the panels show the 
contours satisfying $R_c=4~{h^{-1}\rm Mpc}$ and 
$R_c=6~{h^{-1}\rm Mpc}$, where the characteristic correlation
length $R_c$ is defined by $\xi_0(R_c)=1$.
The observational constraint is not strict at present
because of statistical errors. Then Figure 1 should be regarded as a
demonstration. However, it is instructive. Taking the constraint 
$4~{h^{-1}\rm Mpc}\leq R_c\leq 6~{h^{-1}\rm Mpc}$ into account,
the ratio of the correlation function is 
$\xi_2/\xi_0\simeq-2.0\sim-1.4$ for the $\Lambda$CDM model, and 
$\xi_2/\xi_0\simeq-1.4\sim-1.0$ for the QCDM model. 
We fixed as $R=20h^{-1}{\rm Mpc}$ in Figure~1, however,
similar behavior appears at other values of $R$.
{}For example, in the case $R=30h^{-1}{\rm Mpc}$,  
$\xi_2/\xi_0\simeq-3.5\sim-2.3$ for the $\Lambda$CDM model
and $\xi_2/\xi_0\simeq-2.3\sim-1.8$ for the QCDM model.
The reason why $\xi_2/\xi_0$ depends on the cosmological model 
can be explained as follows. One reason is the difference of the
linear growth rate $D_1(z)$, which affects the Kaiser factor. 
The other effect is the scaling effect due to the geometric 
distortion, which is described by the factors $\cpara$ and $\cperp$.  
The latter effect is dominant.
It is clear from Figure 2 that the less negative value of 
$\wq$ $(\wq>-1)$ decreases the coefficients $\cpara(z)$ 
and $\cperp(z)$. This causes the difference in 
$\xi_2/\xi_0$ depending on the cosmology. 

Now let us discuss a possible effect from stochastic character 
of the bias by considering the case $\callr(z)\neq1$ in 
(\ref{appendixP}). For simplicity we consider the case $\callr(z)$ 
is constant. The panel (c) in Figure 1 plots the contour of 
$\xi_2/\xi_0$ of the $\Lambda$CDM model same as the panel (a) 
but with $\callr(z)=0.6$. 
As is clear from the panels (b) and (c), the both panel becomes
almost same. That is, the ratio $\xi_2/\xi_0$ takes 
the same values in the cases $(\wq=-1/2,~\callr=1)$ and 
$(\wq=-1,~\callr=0.6)$. 
Thus the signal from the cosmological 
model can be confused with that from the stochastic character of the bias.
This fact suggests that an optimistic geometric test suffers 
from degeneracy between the cosmological model parameter and the 
stochasticity parameter of the bias. 

The panel (d) in Figure 1 is same as the panel (b) but with 
$\Omega_m=0.35$, which shows $\Omega_m$-dependence. By comparing
(d) and (b), ambiguity of the  density parameter 
$\Omega_m$ might not be negligible too. 

To solve the degeneracy due to the stochastic character of the bias,
$\xi_4(R)$ might be useful. 
{}Figure 3 plots contours of $\xi_4/\xi_0$ for the various 
cosmological models, whose parameters are same as those in Figure 1.
Similar feature to Figure 1 can be seen in Figure 3. 
However, Figure 3 shows that $\xi_4/\xi_0$ is rather insensitive to 
the stochasticity $\cal R$ and the density parameter $\Omega_m$, 
as is expected from the investigation of the linear stochastic 
biasing in redshift-space (Pen 1998).
Therefore $\xi_4/\xi_0$ might be useful to break the degeneracy,
if it could be measured precisely. However, the amplitude of the 
signal is rather small, which is of order of $10$\% compared 
with $\xi_2/\xi_0$.

\section{Discussions}

In the present paper, we have examined the sensitivity of correlation 
functions in a QCDM cosmological model to the bias evolution, 
incorporating the various observational effect, i.e., the 
light-cone effect, the linear distortion effect, and the non-linear 
and the finger of God effects. Then the feasibility of the geometric 
test is discussed as a probe of the cosmic equation of state assuming 
the future 2dF QSO sample. The amplitude of the correlation functions 
is sensitive to the mean amplitude of the bias, and is rather 
insensitive to the speed of evolution due to the light-cone 
effect. We have found that the systematic difference appears in the 
ratio of the correlation functions depending on the effective cosmic 
equation of state, due to the geometric distortion effect. 
We have also shown that, if the QSO bias has a stochastic
character, the signal from the cosmological model can be confused with 
that from the stochastic bias. Hence the simple geometric test 
with only $\xi_2/\xi_0$ suffers from the degeneracy between the 
cosmological parameter and the bias parameter unless the 
stochasticity character is clarified. For example, 
the $\Lambda$CDM model with $\callr(z)=0.6$ 
and the QCDM model with $\wq=-1/2$ and $\callr(z)=1$ 
predict almost same value of $\xi_2/\xi_0$ at $R=20h^{-1}{\rm Mpc}$.  
Therefore other cosmological information is required, e.g., the higher 
order multipole moment of the correlation function $\xi_4/\xi_0$. 
However the signal from $\xi_4$ seems to become noisy, 
more detailed investigations will be need for the viability.

Finally it will be worth to discuss the robustness of our results for 
several assumptions adopted in the present paper. Another choice of 
the cosmological redshift-space $s(z)$ alters the shape of the 
correlation function (Paper I), hence the predicted values of $\xi_2/\xi_0$ 
will be altered. However, the sensitivity on the cosmic equation of 
state will not be significantly altered, neither will be the feasibility 
of the geometric test. The bias model (equation [7]) seems 
to express general evolution of the bias, then our result is 
not sensitive to the bias model unless the scale-dependece of the 
bias is significant.
Concerning the QCDM cosmological model, our investigation is 
restricted to the case $\wq$ is a constant. And we used the 
fitting formula by Ma et al. (1999), which is applicable to
that case. In most cases, the quintessence equation of state 
changes slowly with time, however, we believe that predictions 
are well-approximated by treating $\wq$ as an averaged constant 
value (e.g., Wang et al. 1998). An open CDM model shows
the similar result with the QCDM model in $\xi_2/\xi_0$ (Paper I), 
then our conclusion is based on the assumption of the spatially 
flat universe.

\vspace{2mm}
We thank Y.~Kojima, A.~Taruya and T.~Chiba for useful discussion and comment.
This work is supported by the Inamori Foundation. 

\vspace{5mm}
\bibliography{bibliography}

%
%
%
\newpage
\begin{figure}[t]
\centerline{\epsfxsize=15cm \epsffile{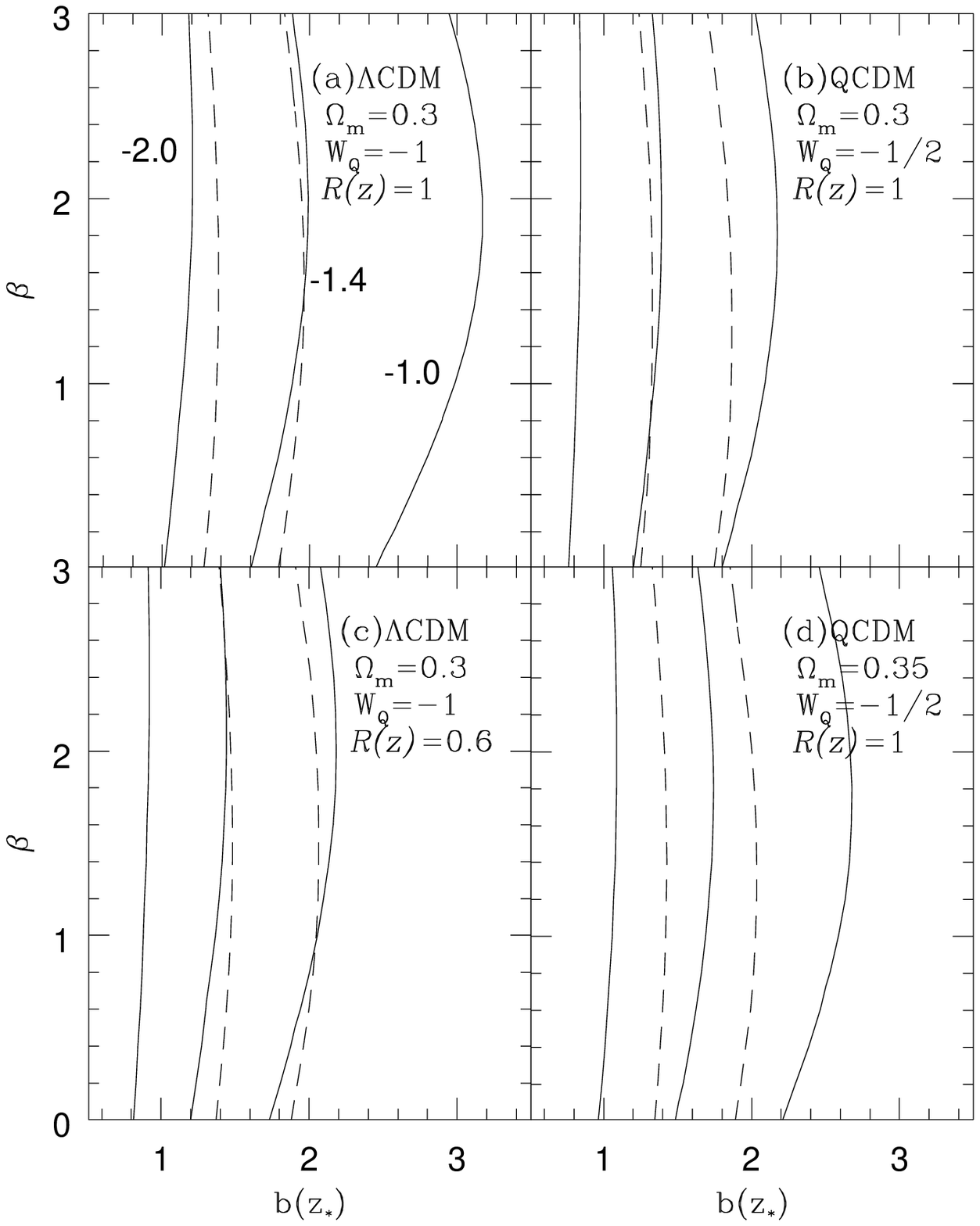}}
\caption{
 ({\it{Solid line}})
Contours of $\xi_2(R)/\xi_0(R)$ at $R=20 h^{-1}{\rm Mpc}$
on the $(\beta-b(z_*))$ plane for various cosmological models: 
(a) $\Lambda$CDM model $(\Omega_m=0.3,\wq=-1.0)$;
(b) QCDM model $(\Omega_m=0.3,\wq=-1/2)$;
(c) same as the panel (a) but with the stochastic bias 
    case with ${\cal R}(z)=0.6$; 
(d) QCDM model $(\Omega_m=0.35,\wq=-1/2)$.
The contour levels of the solid line are indicated on the figure, 
and the same contour levels are adopted for each panel.
In each panel we adopted $h=0.7$, $\Omega_bh^2=0.015$, 
and $\alpha=0.5$ for the bias model. Except for the panel (c), 
the deterministic bias is considered. 
The left (right) {\it dashed line} shows the contour that 
satisfies the condition of the characteristic
correlation length $R_c=4h^{-1}{\rm Mpc}$ ($R_c=6h^{-1}{\rm Mpc}$).
\label{fig1}}
\end{figure}
\begin{figure}[t]
\centerline{\epsfxsize=15cm \epsffile{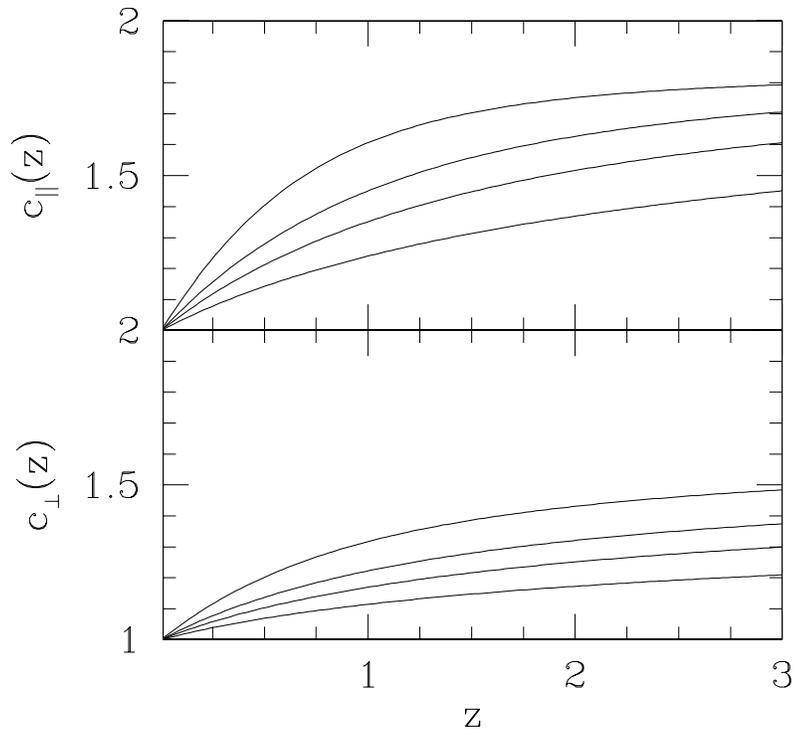}}
\caption{Evolution of $\cpara(z)$ and $\cperp(z)$.
The lines show the cases $\Omega_m=0.3$ and 
$\wq=-1$, $-2/3$, $-1/2$ and $-1/3$, from top to bottom, 
for both panels.
\label{fig2}}
\end{figure}
\begin{figure}[t]
\centerline{\epsfxsize=15cm \epsffile{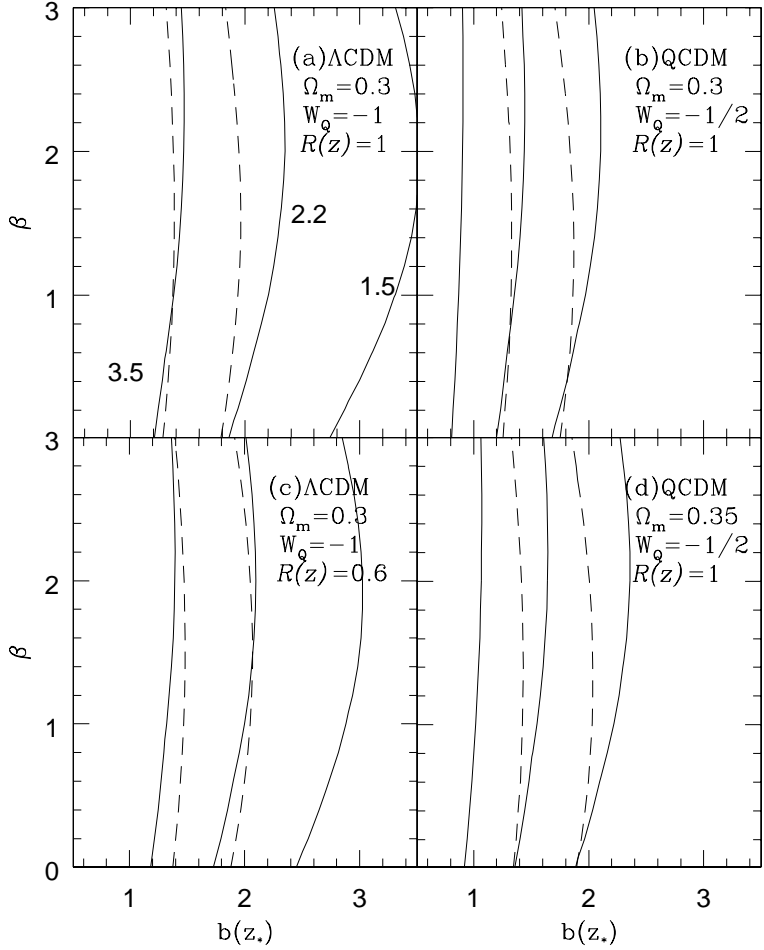}}
\caption{Contours of $\xi_4(R)/\xi_0(R)\times 10$ at 
$R=20 h^{-1}{\rm Mpc}$ on the $(\beta-b(z_*))$ plane for 
various cosmological models.
The model parameters and the meanings of lines are same 
as those of Figure 1. 
\label{fig3}}
\end{figure}
\end{document}